\documentclass[11pt]{article}
\usepackage{hyperref}
\pdfoutput=1
\begin{document}
\title{Remote control of moving sessile droplets}
\author{Xavier Noblin, Richard Kofman and Franck Celestini \\
\\\vspace{6pt} Laboratoire de Physique de la Matiere Condensee,\\
\\\vspace{2pt} CNRS-Universite de Nice Sophia-Antipolis,\\
\\\vspace{2pt} Parc Valrose, 06108 Nice Cedex 2, France \\
\\franck.celestini@unice.fr, xavier.noblin@unice.fr} \maketitle
\begin{abstract}
We recently put in evidence (\cite{noblin2}) that combined vertical and horizontal vibrations can induced a
controlled motion of a sessile supported drop. In this video we generalize this finding and demonstrate that a
remote 2d controlled motion is possible. This video was submitted as part of the Gallery of Fluid Motion 2010
which is showcase of fluid dynamics videos.
\end{abstract}
\section{Introduction}

Microfluidics motivate both fundamental and applied researches (\cite{micro}). The challenge is to find an
optimal process to manipulate small liquid quantities in order to study chemical reactions, biological molecules
and processes, or to perform biomedical tests. A possible way is the sessile drop displacement on surfaces, the
other one consisting in the displacement of liquid in small channels. We recently put in evidence that combined
vertical and horizontal vibrations can be used to reach a controlled droplet motion. In this fluid dynamics
video (which can be seen at the following URL:
\href{http://ecommons.library.cornell.edu/bitstream/1813/8237/2/LIFTED_H2_EMS T_FUEL.mpg}{Video 1}) we
generalize this finding and show that a simple vibrating apparatus can be efficiently used to control the motion
of supported droplets.




\section{Videos description by parts.}

1) Part I. We show that the motion of a supported water droplet is remotely controlled using a joystick. Both
up, down, right and left motions are controlled. The substrate is made of a PDMS elastomer and the water droplet
has an equilibrium wetting angle around $100\,^{o}$.

2) Part II. We are now able to drive the droplet to a target deposited on the substrate. The target is a dye
powder (fluorescein) deposited on the upper right of the substrate. Note that the used of vibrations enhanced
the mixing of the powder within the droplet.

3) Part III. We use a stroboscopic illumination to explain the motion of the drop. Vertical vibrations are
inducing a "pumping mode" (\cite{chodu}, \cite{noblin1}) while horizontal ones are inducing the "rocking mode"
(\cite{celestini}). When both vibrations are applied, a ratchet-like motion is induced (\cite{noblin2}). The
direction of the motion is controlled through the phase shift between the two vibrations. In the present
experiment the horizontal and vertical vibration frequencies are equal to 70 Hz.

4) part IV. We illustrate the ability for the drop to be completely controlled by our joystick. The game for the
big drop, is to eat the other small droplets.


The vibration frequencies are the same in x, y, z directions, and fixed here to 70 Hz. The drop is of order one
millimeter in radius. The liquid is pure water or fluorescein solution. The substrate is made of a piece of PDMS
elastomer.



\end{document}